\documentclass{ctexart}
\usepackage{physics_report}
\usepackage{zhlipsum, lipsum}
\usepackage{cite}
\usepackage{bbm}
\usepackage{mathrsfs}
\usepackage{amsmath}
\usepackage{amsfonts}
\usepackage[colorlinks=true,linkcolor=blue,urlcolor=blue,citecolor=blue,anchorcolor=blue]{hyperref}
\usepackage{graphicx,epstopdf}
\usepackage{subfigure}
\usepackage{epsfig}
\usepackage{dcolumn}
\usepackage{bm}
\usepackage{color}
\usepackage{amssymb}
\usepackage{xcolor}
\usepackage{braket}
\usepackage{ulem}
\usepackage{float}
\usepackage{lipsum}
\usepackage{pifont}
\usepackage{titlesec}
\title[
    Advances in non-Hermitian dynamics of quadratic\superscript{*}
]{
    二次型玻色系统中非厄米动力学的研究进展\superscript{*}
}
\author[
    ZHAO Huawei\superscript{1)} \and LIU Xinlei\superscript{1)} \and HUANG Xinyao\superscript{1)\dagger} \and ZHANG Guofeng\superscript{1)\dagger}
]{
    赵华伟\superscript{1)} \and 刘鑫磊\superscript{1)} \and 黄馨瑶\superscript{1)\dagger} \and 张国锋\superscript{1)\dagger}
}
\institude[
    1) School of Physics, Beihang University, Beijing China
]{
    1)北京航空航天大学 \ 物理学院，北京
}

\begin{document}

    \maketitle
    \abstract{
        非厄米物理是近年来快速发展的重要前沿研究领域, 揭示了诸多新奇物理现象与功能应用. 然而, 当前非厄米物理的研究多集中于经典系统, 非厄米特性对量子效应的影响作为亟待探索的关键科学问题, 已发展为非厄米物理与量子物理交叉领域的新兴研究方向. 无耗散二次型玻色系统中的压缩作用, 既能使系统动力学演化呈现等效非厄米特性, 又能诱导系统产生量子关联效应. 因此, 该系统不仅为实现量子体系中多样化的非厄米动力学提供了天然载体, 更为深入探究非厄米动力学与量子关联等效应的内在联系、实现基于非厄米特性的量子调控提供了重要平台. 本综述将介绍二次型玻色系统中实现非厄米动力学的物理原理, 回顾非厄米动力学诱导的典型物理效应, 并总结基于非厄米动力学调控系统量子关联的近期研究进展.
    }
    \keywords{非厄米动力学, 二次型玻色系统, 非互易, 非厄米趋肤效应, 量子关联}

    \vspace{1em}
    \fund{国家自然科学基金(批准号:12474353, 12474354 )、航空科学基金(批准号:20240058051004)和中央高校基本科研业务费资助的课题．}

    \vspace{1em}
    \email{xinyaohuang@buaa.edu.cn  \quad gf1978zhang@buaa.edu.cn}
    
    \section{引~~~~言}
\label{sec1}
在标准量子力学中, 通常用厄米哈密顿量来描述封闭物理系统的幺正演化. 随着近年来理论和实验的发展, 研究发现利用非厄米哈密顿量可以有效描述诸如存在增益耗散的光学或声学系统、与环境有耦合的开放系统、以及包含相互作用或杂质的凝聚态系统等一系列非厄米物理系统\cite{Bender_2007,RevModPhys.93.015005,Ashida02072020,Zhang31122022}. 由于哈密顿量具备非厄米特性, 这类非厄米系统具有许多不同于厄米系统的新奇性质. 例如, 在一定条件下, 系统满足宇称-时间(Parity-time, PT)对称和反对称性\cite{PhysRevLett.80.5243,nature2012488,science.abf6873}, 存在实能谱, 能够发生PT对称和破缺的相变\cite{WDHeiss_2004}. 相变点出现在本征值和本征态都简并的非厄米奇异点(Non-Hermitian exceptional point, EP), 该点对信号具有非线性响应特性\cite{PhysRevLett.112.203901,Hodaei2017EnhancedSA,Chen2017ExceptionalPE}. 在非厄米系统中还会出现非互易\cite{natphy2023,PhysRevX.8.041031,2024Optomechanical}、非厄米趋肤效应\cite{PhysRevX.8.041031,PhysRevB.106.024301,PhysRevLett.132.096501}、非厄米拓扑相变\cite{PhysRevA.96.032103,PhysRevB.99.081103}和非厄米Aharonov–Bohm(AB)笼\cite{PhysRevA.108.023518,PhysRevResearch.2.033127}等现象. 这些丰富的物理现象不仅有助于我们进一步发展非厄米物理基础理论, 而且在光场调控新方法、新型光学器件设计、测量灵敏度提升等方面具备广阔的应用前景. 因此, 非厄米物理学已发展成为广受关注的研究方向, 引起了人们的极大兴趣. 传统的非厄米系统通过在模型中引入非对称耦合(绝对值不相等的非对角矩阵元)\cite{PhysRevLett.126.216405,PhysRevB.106.035425,PhysRevB.105.205402}、在位能引入对角增益或耗散(对角矩阵元有虚部)\cite{Hodaei2017EnhancedSA,PhysRevLett.123.170401}或者耗散耦合(非对角矩阵元的相位不满足复共轭关系)\cite{PhysRevLett.124.030401,PhysRevLett.133.133601}来实现, 上述方法普遍都是从系统出发构造非厄米哈密顿量, 使得系统演化出现非厄米特性.

然而, 非厄米特性并非只出现在非厄米哈密顿量中. 人们在二次型费米模型的研究中发现, 除了在位能量项$a_j^{\dagger}a_j$和跃迁相互作用项$a_j^{\dagger}a_k$, 该模型的哈密顿量中还会存在由成对产生或湮灭算符构成的配对项$a_j^{\dagger}a_k^{\dagger}$和$a_j a_k(j\neq k)$. 通过将哈密顿量嵌入以$( a_1,..., a_j, a_1^\dagger,..., a_j^\dagger)^\text{T}$作为实空间基底向量或以$( a_k, a_{-k}^\dagger)^\text{T}$作为动量空间基底向量的Bogoliubov–de Gennes (BdG)框架中\cite{PhysRevLett.100.200404,PhysRevLett.110.131601}, 可以发现此时虽然系统具有厄米的二次型哈密顿量, 但配对项的出现使其在BdG框架下的动力学行为出现等效的非厄米特性. 例如基于最近邻跃迁和配对项构成的一维费米Kiteav链\cite{2001Kiteav,PhysRevB.103.134507,PhysRevB.105.024514,PhysRevResearch.5.L012012}由厄米哈密顿量描述,却同时在BdG框架下具有非厄米动力学演化矩阵, 从而引发非厄米Majorana零模\cite{PhysRevB.100.085110,PhysRevB.104.205131}, 非厄米趋肤效应\cite{PhysRevLett.124.086801,PhysRevB.108.195126}, 非布洛赫拓扑相变\cite{PhysRevLett.118.267701,PhysRevB.101.195147,PhysRevB.109.115115,PhysRevLett.127.270602}和非阿贝尔统计\cite{PhysRevLett.96.016803}等. 这些研究进展奠定了“非厄米拓扑超导”的理论与实验基础, 已经发展为相对成熟的研究方向.

\begin{figure}[H]
\vspace*{2mm}\centering
		\includegraphics[angle=0,width=1\linewidth]{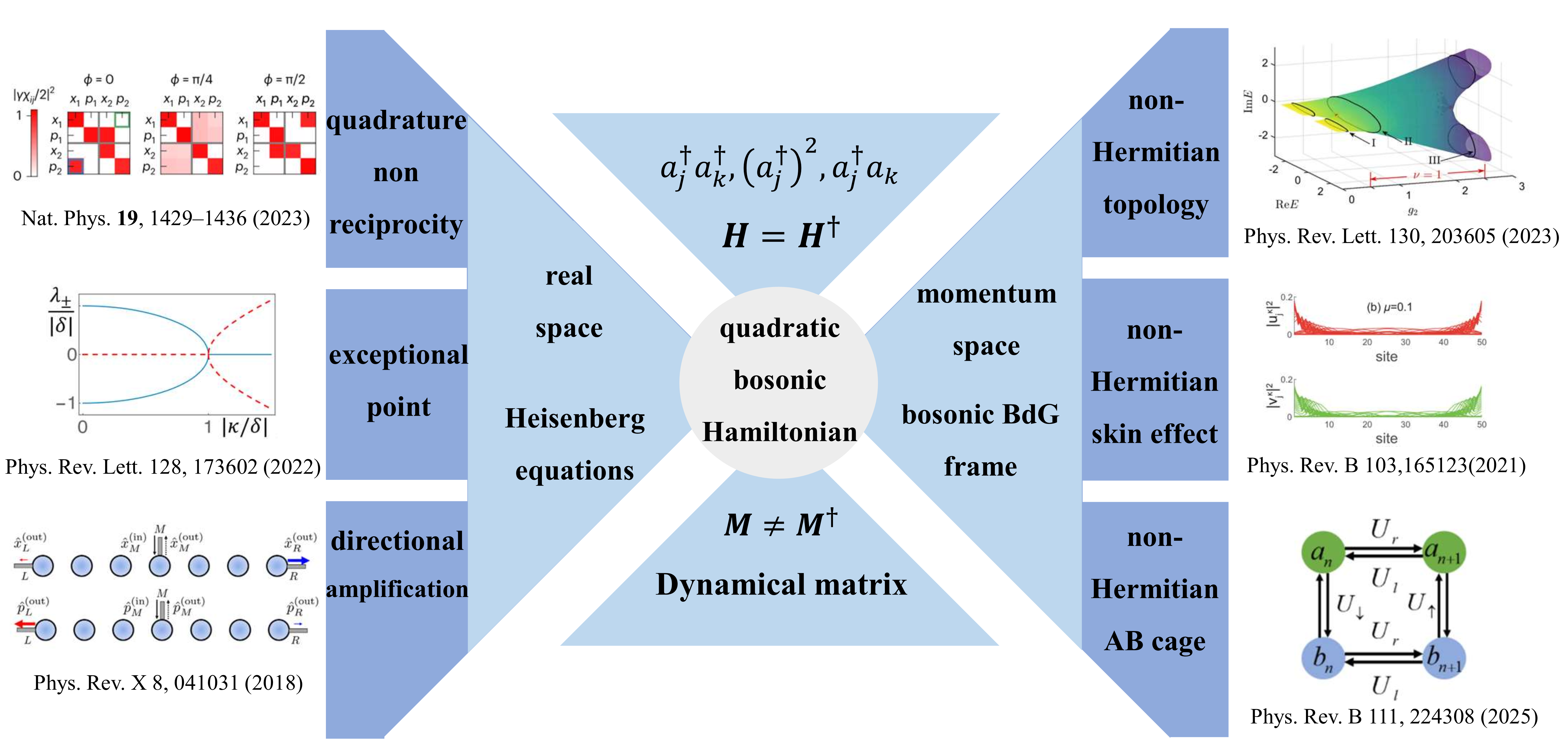}
\bicaption{由厄米哈密顿量描述的QBS在实空间和动量空间下具有等效的非厄米动力学演化矩阵,可诱导出正交性非互易、EP、手性传输,以及点隙拓扑相变、非厄米趋肤效应和非厄米AB笼等新奇物理现象.}
{The QBS described by a Hermitian Hamiltonian exhibits equivalent non-Hermitian dynamical evolution matrices in both real and momentum space, which can induce novel physical phenomena such as quadrature nonreciprocity, EP, chiral transport, as well as point-gap topological phase transitions, the non-Hermitian skin effect, and non-Hermitian Aharonov–Bohm cages. }
\vskip 1mm
\end{figure}
近几年研究者开始将上述二次型费米模型映射到二次型玻色系统(quadratic bosonic system, QBS)中. 2018年Clerk等\cite{PhysRevX.8.041031}研究了玻色Kiteav链(Bosonic Kitaev chain, BKC): 在一维玻色链上, 相邻玻色模式间的分束器型(beamsplitter, BS)相互作用$a_j^{\dagger}a_{j+1}$和双模压缩(two-mode-squeezing, TMS)相互作用$a_j^{\dagger}a_{j+1}^\dagger$对应费米模型中的跃迁项和配对项, 揭示了二次型玻色哈密顿量中TMS作用诱导的等效非厄米动力学\cite{PhysRevA.88.063631,PhysRevLett.115.245302,PhysRevLett.117.045302,PhysRevX.6.041026,Flynn_2020}. 之后, 基于两体QBS\cite{PhysRevA.99.063834,PhysRevResearch.6.043209,natphy2023,PhysRevLett.128.173602}以及BKC\cite{PhysRevB.103.165123,y174pms8,PhysRevX.8.041031,PhysRevA.100.062323,PhysRevLett.127.245701,PhysRevB.105.085423,PhysRevA.106.053315,PhysRevLett.131.053001,2024Optomechanical,2024Quantum,PhysRevB.111.035131}、玻色Su-Schrieffer-Heeger(SSH)模型\cite{PhysRevLett.130.203605,PhysRevB.106.024301}对应的多体QBS, 人们也发现了诸多基于非厄米动力学的新奇物理效应\cite{natphy2023,PhysRevLett.128.173602,y174pms8}. 例如在两体系统和BKC模型中, 通过正交场算符变换得到正交非互易和相位可控的手性传输以及传输信号的缩放\cite{natphy2023,PhysRevX.8.041031,2024Optomechanical}, 为量子线路和片上集成等量子信息技术提供了新思路; 在BKC模型和引入压缩项的玻色SSH模型中, 由近邻TMS诱导的非厄米动力学可用于实现双边趋肤效应\cite{PhysRevX.8.041031,PhysRevB.106.024301,PhysRevLett.132.096501}、边界敏感放大和拓扑相变\cite{PhysRevB.103.214306,PhysRevLett.130.203605}等. 进一步的, QBS本身作为量子系统, 压缩作用的存在会诱导系统中产生量子关联等量子效应. 有别于在传统开放量子系统中基于量子轨迹等方法消除量子跳跃效应的影响构造等效非厄米哈密顿量\cite{science.1258004,PhysRevLett.113.053604}以及保留量子跳跃效应基于刘维尔谱来实现非厄米特性的方式\cite{PhysRevLett.116.240404,PhysRevA.98.042118}, 含有压缩作用的QBS作为无需引入耗散或增益的量子平台能够免受量子噪声或跳跃的影响, 为研究非厄米动力学与量子效应的内在联系提供了天然载体\cite{PhysRevResearch.7.L022034,quanEP1,quanEP2}, 如近年来人们发现基于动力学EP诱导系统中压缩和纠缠的动力学行为突变\cite{PhysRevLett.128.173602,yu2025}、增强量子Fisher信息上限\cite{fisherEP},以及由拓扑相诱导产生稳态纠缠及相变\cite{PhysRevA.108.062405,PRXQuantum.5.010313}等. 上述进展不仅对玻色系统的非厄米动力学图像做出了极大的丰富, 也为量子传感、非互易器件、纠缠调控和拓扑量子计算等领域提供了新的思路和实验平台.
 
\section{二次型玻色系统}
\label{sec2}
玻色子是自旋量子数为整数倍的粒子,其产生湮灭算符满足对易关系$[a_j,a_k^\dagger]=\delta_{jk}$. 光子、磁振子、声子、极化激元和冷原子等均是现代物理研究中的常见玻色子, 光学微腔、波导腔和声子-极化激元腔等也是典型的玻色场系统. 因此玻色子在光与物质相互作用、量子计算和处理、量子通信、冷原子研究和量子超流等诸多领域都有广泛应用. 由于玻色系统不受泡利不相容原理的限制, 其还能在单格点处出现费米系统中被禁止存在的单模压缩(single-mode-squeezing, SMS)项$(a_j^\dagger)^2$. QBS的普适哈密顿量有如下表达:
\begin{equation}\label{generalQBS}
\begin{split}
    H_{\text{QBS}}=\sum_{j,k}^N \left(J e^{i \phi_J}a_{j}^\dagger a_k +\kappa e^{i \phi_{\kappa}} a_{j}^\dagger a_k^{\dagger}\right) +\sum_j^N \eta (a_j^\dagger)^2+\mathrm{H.c.},
    \end{split}
\end{equation}
其中$a_{j}^\dagger a_k,a_{j}^\dagger a_k^\dagger(j\neq k)$分别代表任意两个不同模式之间的BS相互作用(强度$J$, 可调相位$\phi_{J}$)和TMS相互作用(强度$\kappa$, 可调相位$\phi_{\kappa}$), $(a_j^\dagger)^2$则代表每个模式的SMS(强度$\eta$), 此处省略所有模式自身的在位能项$a_{j}^\dagger a_j$, 其对体系的结构与能谱特征等不产生影响. 通过玻色BdG框架\cite{COLPA1978327,PhysRevB.87.174427,PhysRevB.89.054420,PhysRevB.98.115135,PhysRevA.101.013625,PhysRevA.103.L051301,PhysRevA.104.013305,PhysRevB.103.165123,y174pms8,PhysRevB.103.214306,PhysRevB.105.224301,PhysRevB.111.L241401}和海森堡动力学方程\cite{natphy2023,PhysRevLett.128.173602,PhysRevX.8.041031,2024Optomechanical}$i\partial_t \Psi=M\Psi (\hbar=1)$得到实空间下QBS的非厄米动力学演化矩阵有如下区块化特征:
\begin{equation}\label{BdGreal}
\begin{split}
    M =
\begin{pmatrix}
\boldsymbol{\xi} & \boldsymbol{\Delta}\\[4pt]
\boldsymbol{-\Delta}^\dagger & \boldsymbol{-\xi}^\dagger
\end{pmatrix}_{\!2N\times 2N},
    \end{split}
\end{equation}
其中$\Psi = (a_1,\dots,a_N, a_1^\dagger,\dots,a_N^\dagger)^\text{T}$是BdG框架下的$2N$维实空间基底向量, $N$维区块厄米矩阵$\boldsymbol{\xi}=\boldsymbol{\xi}^\dagger$代表哈密顿量中的BS相互作用项和玻色子的自身能量项, 区块对称矩阵$\boldsymbol{\Delta}=\boldsymbol{\Delta}^\dagger$中包含TMS和SMS相互作用项.

在对QBS进行实空间到动量空间($q$)的投射过程中需要对每个格点的产生湮灭算符做傅里叶变换$a_k = \frac{1}{\sqrt{N}}\sum_q e^{iqk}a_q,
a_k^\dagger = \frac{1}{\sqrt{N}} \sum_{q}e^{-i q k} a_q^\dagger $, 其中$ q = 2\pi n/N , n = 0, 1, 2, \ldots, N-1 $, 并经过归一化公式$\frac{1}{N}\sum_{k} e^{i (q' - q) k} =\delta_{qq'}, \frac{1}{N}\sum_{q} e^{i (k' - k) q} =\delta_{kk'}$后, 动量空间的二次型哈密顿量在BdG框架下有$H(q)=\frac{1}{2}\Psi^\dagger(q) H_{\text{BdG}}(q)\Psi(q)$, 而动量空间动力学演化矩阵$ M(q)$为:
\begin{equation}\label{BdGq}
\begin{split}
   M(q)=\sigma_z H_{\text{BdG}}(q) =
\begin{pmatrix}
\xi(q)& \Delta(q)\\[4pt]
-\Delta^*(-q) & -\xi^*(-q)
\end{pmatrix},
    \end{split}
\end{equation}
其中$\Psi(q) = (a_q,a_{-q}^\dagger)^\text{T}$是BdG框架下的$2$维动量空间基底向量, 负动量$(-q)$是为了满足体系动量守恒而自然出现的结果, $H_{\text{BdG}}(q)$为动量空间哈密顿矩阵, $\sigma_z$为泡利$z$算符. $\xi(q), \xi^*(-q)$代表BS相互作用项和玻色子能量项, $\Delta(q), \Delta^*(-q)$中包含SMS和TMS相互作用项. 

\begin{figure}[H]
\vspace*{2mm}\centering
		\includegraphics[angle=0,width=0.5\linewidth]{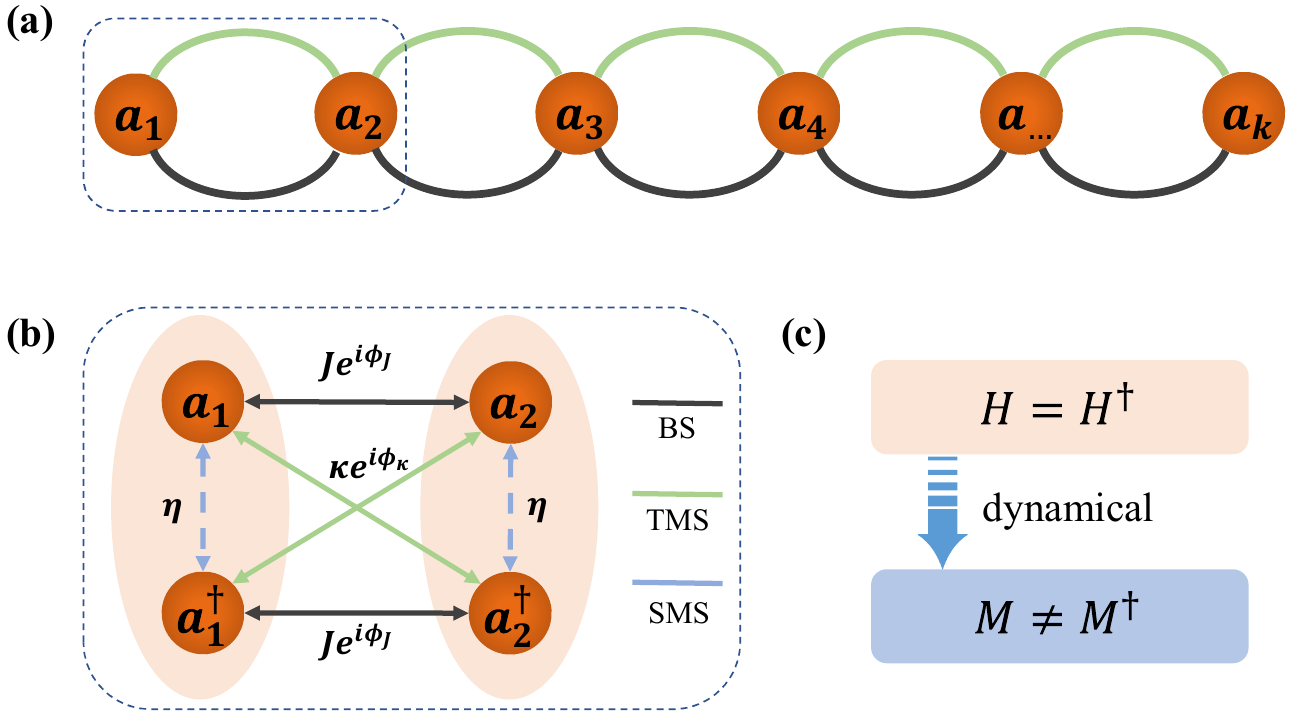}
\bicaption{(a) BKC模型示意图. (b) 两模QBS示意图. 两个玻色模式由强度为$J$的BS相互作用和强度为$\kappa$的TMS相互作用耦合, 分别伴随可调耦合相位$\phi_J$和$\phi_\kappa$, BKC中SMS的强度为$\eta=0$. (c) QBS具有厄米哈密顿量$H$和BdG框架下的非厄米动力学演化矩阵$M$.}
{(a) Schematic of BKC. (b) Schematic of a two-mode QBS.  The two bosonic modes are coupled by a BS interaction of strength $J$ and a TMS interaction of strength $\kappa$, both accompanied by tunable coupling phases $\phi_J$ and $\phi_\kappa$. In BKC, the strength of SMS is $\eta=0$. (c) The Hamiltonian $H$ of the QBS is Hermitian, whereas its dynamical evolution matrix $M$ under BdG framework is non-Hermitian.}\label{model}
\vskip 1mm
\end{figure}

在规定QBS即方程(\ref{generalQBS})中的SMS系数$\eta=0$且保留最近邻相互作用$k=j+1$后得到如图\ref{model}(a)所示的一维BKC模型, 其哈密顿量为:
\begin{equation}\label{BKCphi}
\begin{split}
    H_{\mathrm{BKC}}=\sum_{j}^N \left(J e^{i \phi_J}a_{j}^\dagger a_{j+1} +\kappa e^{i \phi_{\kappa}} a_{j}^\dagger a_{j+1}^{\dagger}+\mathrm{H.c.}\right) ,
    \end{split}
\end{equation}
显然该哈密顿量本身满足厄米关系$H_{\mathrm{BKC}}^\dagger=H_{\mathrm{BKC}}$. 当同种相互作用取相同系数, BKC可以看作是如图\ref{model}(b)所示的两体二次型玻色单元在一维的拓展, 其厄米哈密顿量为$H=Je^{i\phi_J}a_1^\dagger a_2 +\kappa e^{i\phi_{\kappa}} a_1^\dagger a_2^\dagger +\mathrm{H.c.}$ 在BdG框架下通过海森堡方程$i\partial_t ( a_1, a_2, a_1^\dagger, a_2^\dagger)^\text{T}=M( a_1, a_2, a_1^\dagger, a_2^\dagger)^\text{T} (\hbar=1)$对该体系作动力学描述得到其动力学演化矩阵:
\begin{equation}\label{n=2juzhen}
	\begin{split}
	M=
	\begin{pmatrix}
		0&Je^{i\phi_J}&0&\kappa e^{i\phi_\kappa}\\
		Je^{-i\phi_J}&0&\kappa e^{i\phi_\kappa}&0\\
		0&-\kappa e^{-i\phi_\kappa}&0&-Je^{-i\phi_J}\\
		-\kappa e^{-i\phi_\kappa}&0&-Je^{i\phi_J}&0\\
	\end{pmatrix}.
	\end{split}
\end{equation}
与系统自身的哈密顿量$H_{\mathrm{}}$相比, 其动力学演化矩阵$M$在次对角线上的矩阵元不再互为复共轭, 导致厄米哈密顿量中出现非厄米动力学演化矩阵$M^\dagger\neq M$. 

对格点数为$N$的BKC模型则以$2N$维的列向量$\Psi=( a_1,..., a_N, a_1^\dagger,..., a_N^\dagger)^\text{T}$作为基底进行描述. 通过海森堡方程$i\partial_t \Psi=M_{\mathrm{BKC}}\Psi (\hbar=1)$得到关于BKC的$2N$维非厄米动力学演化矩阵:
\begin{equation}\label{BKCjuzhen}
\begin{split}
M_{\text{BKC}} &=
\left[\begin{array}{c|c}
\phantom{-}J\!\bigl(e^{-i\phi_J}\mathcal{M}_{\downarrow}+e^{i\phi_J}\mathcal{M}_{\uparrow}\bigr)
& \kappa e^{i\phi_{\kappa}}\bigl(\mathcal{M}_{\downarrow}+\mathcal{M}_{\uparrow}\bigr)\\[4pt]
\hline
-\kappa e^{-i\phi_{\kappa}}\bigl(\mathcal{M}_{\downarrow}+\mathcal{M}_{\uparrow}\bigr)
& -J\!\bigl(e^{i\phi_J}\mathcal{M}_{\downarrow}+e^{-i\phi_J}\mathcal{M}_{\uparrow}\bigr)
\end{array}\right]_{2N\times 2N},
\end{split}
\end{equation}
其中
\begin{equation}
\begin{split}
\mathcal{M}_{\downarrow} &=
\begin{bmatrix}
0 & 0 & 0 & \cdots & 0 \\
1 & 0 & 0 & \cdots & 0 \\
0 & 1 & 0 & \cdots & 0 \\
\vdots & \vdots & \ddots & \ddots & \vdots \\
0 & 0 & \cdots & 1 & 0
\end{bmatrix}_{N\times N},
\quad
\mathcal{M}_{\uparrow} =
\begin{bmatrix}
0 & 1 & 0 & \cdots & 0 \\
0 & 0 & 1 & \cdots & 0 \\
\vdots & \vdots & \ddots & \ddots & \vdots \\
0 & 0 & \cdots & 0 & 1 \\
0 & 0 & \cdots & 0 & 0
\end{bmatrix}_{N\times N}.
\end{split}
\end{equation}
$M_{\text{BKC}}$是由四个$N$维矩阵构成的分块矩阵, 主对角线上代表BS相互作用的矩阵块为厄米矩阵, 而次对角线上代表TMS相互作用的两个矩阵块不再互为复共轭, 因此BKC动力学演化矩阵为非厄米矩阵$M_{\mathrm{BKC}}^\dagger\neq M_{\mathrm{BKC}}$.

\begin{figure}[H]
\vspace*{2mm}\centering
		\includegraphics[angle=0,width=0.5\linewidth]{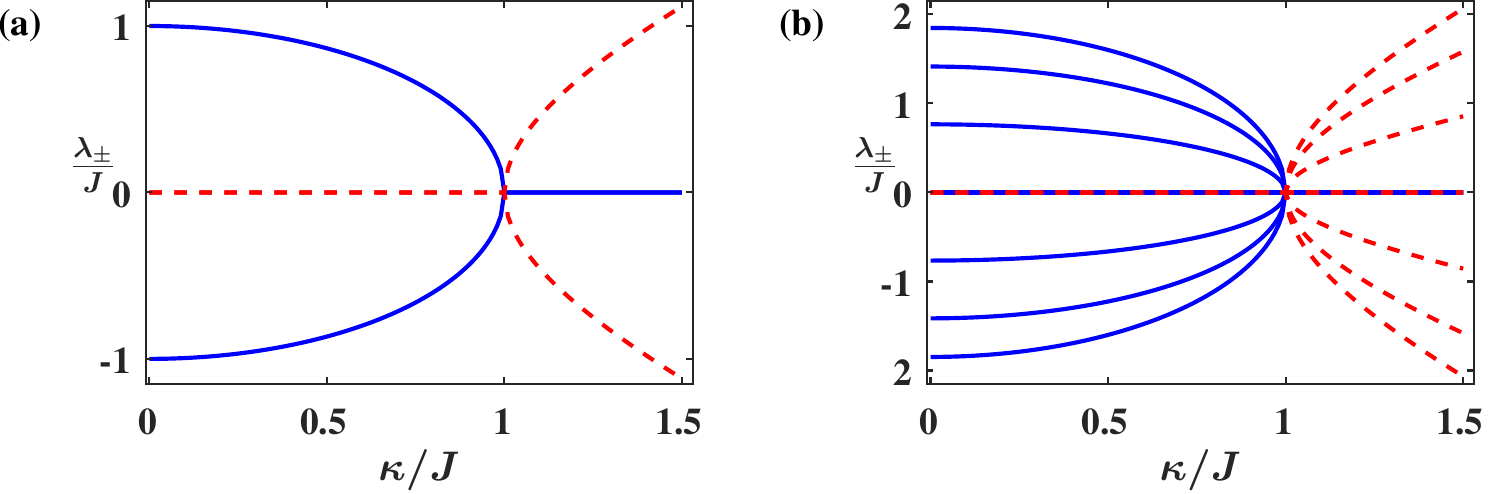}
\bicaption{(a) $N=2$和(b)$N=7$的BKC动力学演化矩阵本征值$\lambda_{\pm}$随耦合系数$\kappa/J$的分布. 蓝色实线代表本征值的实部$\mathrm{Re}(\lambda_{\pm})$, 红色虚线代表本征值的虚部$\mathrm{Im}(\lambda_{\pm})$.}
{Distribution of eigenvalues $\lambda_{\pm}$ for a BKC with (a) $N=2$ and (b)$N=7$ as functions of the coupling coefficient $\kappa/J$. The blue solid line indicates the real part $\mathrm{Re}(\lambda_{\pm})$, and the red dashed line indicates the imaginary part $\mathrm{Im}(\lambda_{\pm})$.}\label{eigen}
\vskip 1mm
\end{figure}

在厄米系统中, 系统的本征值一般为纯实数且本征态之间线性无关. 然而在非厄米系统中, 本征值往往是复数且本征态之间线性相关. 非厄米系统的本征值与本征态发生简并的地方即EP, 这也是独属于非厄米系统的重要特征. 对于两体BKC单元, 通过求解其动力学演化矩阵(\ref{n=2juzhen})的特征方程$\det(M-\lambda I)=0$, 得到两个互为相反数且二重简并、与相位$\phi_J, \phi_\kappa$无关的动力学矩阵本征值$ \lambda_\pm =\pm\sqrt{J^2-\kappa^2}.$ 如图\ref{eigen}(a), 在$|J|<|\kappa|$的情况下, $M$具有两个纯虚数本征值$ \lambda_\pm =\pm i\sqrt{\kappa^2-J^2}$. 与之相反, 在$|J|>|\kappa|$的情况下$M$具有两个纯实数本征值.  在两个区域的分界点$|J|=|\kappa|$处本证能谱完全重叠, 即为系统动力学演化矩阵的二阶EP. 

对比动力学矩阵(\ref{n=2juzhen})和(\ref{BKCjuzhen}), 实际上这两种矩阵都保持了完全相同的区块化特征, 其满足粒子空穴对称性$\mathcal{C}M\mathcal{C}^{-1}=-M$, $\mathcal{C}=\tau_xK$为反幺正算符, 其中$\tau_x=\sigma_x\otimes I_N$, $\sigma_x$为泡利$x$矩阵, $I_N$为$N$阶单位矩阵, $K$为复共轭操作. 粒子空穴对称性会导致矩阵出现关于零能对称的正负本征值, $2N$维非厄米动力学矩阵则会具有$N$个二重简并的本征值. 如图~\ref{eigen}(b), 以$N=7$为例, 其动力学演化矩阵具有$7$个二重简并的本征值, 除去一对本征值为0其余六对本征值都关于$\lambda=0$对称: $ \lambda_\pm =\pm A(j)\sqrt{J^2-\kappa^2}$, 其中$A(j)$是关于格点数$j$的常系数. 与$N=2$的情况相同, 这些本征值在$|J|<|\kappa|$的情况下为纯虚数, 而在$|J|>|\kappa|$的情况下本征值由纯虚数转化为纯实数. 同样的, 所有本征值在$|J|=|\kappa|$点处重叠, 即动力学演化矩阵的高阶EP. 显然, 对于N维BKC模型, 其动力学演化矩阵总是在$|J|=|\kappa|$处出现一个$N$阶EP.

\section{非互易}
\label{sec3}
洛伦兹互易性是在电磁学、光学等领域已经被广泛认知的普适性质. 在电磁学中互易性被定义为在只含一个电压源(电流源)的线性电阻电路中, 电压源(或电流源)与电流表(电压表)互换位置, 电流表(电压表)读数不变. 光学非互易指的是在光波传播过程中交换光源与探测器的位置, 探测结果不变, 即光路传播的可逆性. 与之相反, 非互易性是指波信号或能量等的传输受到传播方向的影响, 即在相同传播路径上的正向和反向传输具有不同的传输效率\cite{RevModPhys.21.463,RevModPhys.29.651}. 判断一个系统互易与否, 最直接的方法是通过分析输入与输出信号之间的散射矩阵是否为对称矩阵.基于洛伦兹互易定理的成立条件,非互易的常见实现方法也可以归为打破对称性\cite{PhysRevLett.105.126804}、打破线性\cite{PhysRevLett.117.123902}和打破非时变性\cite{PhysRevLett.109.033901}等三类物理机制. 

\begin{figure}[H]
\vspace*{2mm}\centering
		 \includegraphics[angle=0,width=0.4\linewidth]{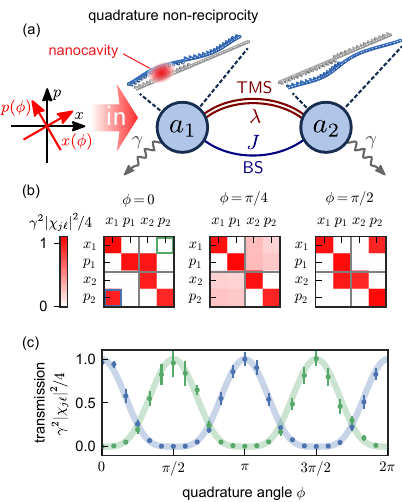}
\bicaption{(a) 正交非互易实验示意图. 左侧为沿相对参考规范(黑色)旋转角度$\phi$的正交分量(红色)施加的相干激发. (b) 正交非互易. 实验测得的极化率矩阵振幅随相位角$\phi$的演化非互易体现在极化率矩阵的非对称特征, 即$|\chi|\neq|\chi|^{\mathrm{T}}$. (c) 绿色($p_2\to x_1$)与蓝色($x_1\to p_2$)框内的矩阵元行为表明入射正交分量角度$\phi$是否可调控非互易. 两机械模经等强度的BS与TMS相互作用($J=\lambda=\gamma/4$), 系统呈现正交非互易传输. 当$\phi=\pi/4$时, 非互易性消失; 当$\phi=\pi/2$时, 传输方向反向\cite{natphy2023}.}
{(a) Experimental schematic of quadrature nonreciprocity. Leftmost: coherent excitation along quadratures (red) rotated by $\phi$ from the reference gauge (black). (b) Quadrature nonreciprocity, which corresponding measured susceptibility matrix amplitudes evolve differently with $\phi$. Nonreciprocity manifests in an asymmetric modulus of the susceptibility $|\chi|\neq|\chi|^{\mathrm{T}}$. (c) The behaviour of the elements within the green and blue boxes shows whether nonreciprocity can be tuned by the incoming quadrature angle. Two modes interact through BS and TMS of equal strength ($J=\lambda=\gamma/4$), showing quadrature nonreciprocity. Nonreciprocity vanishes for $\phi=\pi/4$. Furthermore, setting $\phi=\pi/2$ inverts the transmission direction\cite{natphy2023}.}\label{qdnr}
\vskip 1mm
\end{figure}

2023年, Andreas等\cite{natphy2023}在不打破系统时间反演对称性的前提下, 于纳米腔光力系统中采用正交基变换$x_j=(a_j+a_j^\dagger)/\sqrt{2}$和$p_j=-i(a_j-a_j^\dagger)/\sqrt{2}$实现了$x$和$p$算符之间的正交性非互易传输. 如图\ref{qdnr}(a)所示为两个机械模$a_1$和$a_2$通过光力耦合BS相互作用(强度$J$)与TMS相互作用(强度$\lambda$)进行耦合. 在机械模频率的旋转框架下, 该系统的有效哈密顿量为$ H=Ja_1^\dagger a_2+\lambda a_1^\dagger a_2^\dagger+\mathrm{H.c.}$ 系统的动力学演化通过朗之万方程$\dot{q}=Mq-\sqrt{\gamma}q_{\mathrm{in}}$得到, 其中$M$是动力学演化矩阵, $\gamma = 2π × 3.7 \mathrm{kHz}$是两个机械模的衰减率. 在这项工作中, 基底向量$q$并不是$( a_1, a_2, a_1^\dagger, a_2^\dagger)^\text{T}$, 而是经过正交场算符变换后的新基底向量$( x_1, p_1, x_2, p_2)^\text{T}$. 通过频域下的输入输出理论$q_{\mathrm{out}}=q_{\mathrm{in}}+\sqrt{\gamma}q$\cite{RevModPhys.82.1155}得到的散射矩阵$S(\omega)=I+\gamma(i\omega I+M)^{-1}=I+\gamma\chi(\omega)$是对系统传输特性的解析体现, 其中极化率矩阵$\chi(\omega)=(i\omega I+M)^{-1}$同样反映系统的传输特性. 若$S(\omega)$或$\chi(\omega)$为对称矩阵则互易传输, 反之非对称矩阵元代表非互易传输. 在经过以上正交场算符变换和动力学分析后, 两模BKC的正交算符动力学方程为:
\begin{equation}\label{xpdyna}
\begin{split}
    \dot{x}_1=-\frac{\gamma}{2}x_1 + (J-\lambda)p_2-\sqrt{\gamma}x_{1,\mathrm{in}},\\
     \dot{p}_1=-\frac{\gamma}{2}p_1 -(J+\lambda)x_2-\sqrt{\gamma}p_{1,\mathrm{in}},\\
      \dot{x}_2=-\frac{\gamma}{2}x_2 + (J-\lambda)p_1-\sqrt{\gamma}x_{2,\mathrm{in}},\\
       \dot{p}_2=-\frac{\gamma}{2}p_2 - (J+\lambda)x_1-\sqrt{\gamma}p_{2,\mathrm{in}}.\\
    \end{split}
\end{equation}
在$J=\lambda$下, $ \dot{x}_1$与$p_2$之间发生了解耦, 而$\dot{p}_2$与$x_1$仍然耦合, $x_2$与$p_1$之间同样如此. 这种情况在极化率矩阵中展现出来[图\ref{qdnr}(b)]:
\begin{equation}\label{xpjuzhen}
	\begin{split}
	\chi(\omega=0)=
	\begin{pmatrix}
		-\frac{\gamma}{2}&0&0&0\\
		0&-\frac{\gamma}{2}&\frac{8J}{\gamma^2}&0\\
		0&0&-\frac{\gamma}{2}&0\\
		\frac{8J}{\gamma^2}&0&0&-\frac{\gamma}{2}\\
	\end{pmatrix}.
	\end{split}
\end{equation}
极化率矩阵在次对角元上的非对称项意味着非互易传输的存在. 当一个输入信号被编码在正交算符$x_1$上, 该信号从模$a_1$传输到$a_2$并且在算符$p_2$产生响应: $S_{x_1\to p_2}=\gamma \chi_{41}=8J/\gamma\neq0$. 然而在反向过程中, $x_1$却不能对来自$p_2$的输入信号产生响应: $S_{p_2\to x_1}=\gamma \chi_{14}=0$. 在$x_2$和$p_1$之间发生同样的情况:$S_{x_2\to p_1}\neq0,S_{p_1\to x_2}=0$. 除了在不同正交算符中形成单向非互易传输, 当耦合强度与机械模衰减率之间的强度关系满足$J=\lambda>\gamma/8$时, $S_{x_1\to p_2(x_2\to p_1)}=\gamma \chi_{41(14)}>1$, 反之则有$S_{x_1\to p_2(x_2\to p_1)}<1$, 这意味着该系统还能够实现单向的信号放大或抑制器.

正交非互易对正交变换的相位有灵敏响应, 通过相位调控可以对非互易实现控制. 当正交算符$\{x_j,p_j\}$做旋转角为$\phi$的旋转操作后:
\begin{equation}\label{rotphi}
\begin{split}
    \begin{pmatrix}
		x_j(\phi_j)\\
		p_j(\phi_j)\\
	\end{pmatrix}
    =\begin{pmatrix}
		\cos \phi_j&\sin \phi_j\\
		-\sin \phi_j&\cos \phi_j\\
	\end{pmatrix}
    \begin{pmatrix}
		x_j(0)\\
		p_j(0)\\
	\end{pmatrix}
    =R(\phi_j)
    \begin{pmatrix}
		x_j(0)\\
		p_j(0)\\
	\end{pmatrix},
    \end{split}
\end{equation}
系统极化率矩阵$\chi(J=\lambda)$对应变换为:
\begin{equation}\label{rotchi}
\begin{split}
    \chi_{\phi}&=[\oplus_j^NR(\phi_j)]\chi[\oplus_j^NR(\phi_j)]^\text{T}\\
	&=\begin{pmatrix}
		-\frac{\gamma}{2}&0&-\frac{4J\sin(2\phi)}{\gamma^2}&-\frac{8J\sin^2(\phi)}{\gamma^2}\\
		0&-\frac{\gamma}{2}&\frac{8J\cos^2(\phi)}{\gamma^2}&\frac{4J\sin(2\phi)}{\gamma^2}\\
		-\frac{4J\sin(2\phi)}{\gamma^2}&-\frac{8J\sin^2(\phi)}{\gamma^2}&-\frac{\gamma}{2}&0\\
		\frac{8J\cos^2(\phi)}{\gamma^2}&\frac{4J\sin(2\phi)}{\gamma^2}&0&-\frac{\gamma}{2}\\
	\end{pmatrix}.
    \end{split}
\end{equation}
如图\ref{qdnr}(c)所示, 仍然以$x_1$和$p_2$之间的传输效率$\chi_{x_1\to p_2}=8J\cos^2(\phi)/\gamma,
    \chi_{p_2\to x_1}=-8J\sin^2(\phi)/\gamma$为例: 当$\phi=0$时, 旋转后的极化率矩阵(\ref{rotchi})退化为原始矩阵(\ref{xpjuzhen})并伴随从$x_1\to p_2$的单向传输. 当$\phi=\pi/4$时, 非互易性完全消失: $\chi_{x_1\to p_2}=
    \chi_{p_2\to x_1}=4J/\gamma$. 当$\phi=\pi/2$时, 系统产生了与$\phi=0$时传输方向完全相反的单向非互易$p_2\to x_1$. 

以上结果是利用光力腔系统在实验上对两体模型中场算符正交非互易的实现. Clerk等\cite{PhysRevX.8.041031}研究了近邻纯虚数耦合的BKC模型$H_{\mathrm{B}}=\sum_j (it a_{j+1}^\dagger a_j +i\Delta a_{j+1}^\dagger a_j^{\dagger} +\mathrm{H.c.})$. 以正交场算符$x_j=(a_j+a_j^\dagger)/\sqrt{2}$和$p_j=-i(a_j-a_j^\dagger)/\sqrt{2}$进行变换得到等效哈密顿量为:
\begin{equation}\label{BKCxp}
\begin{split}
    H_{\mathrm{B}}\equiv\sum_j \left[-(t-\Delta) x_{j+1} p_j +(t+\Delta) p_{j+1} x_j \right].
    \end{split}
\end{equation}
系统正交算符的动力学演化可以通过海森堡方程得到:
\begin{equation}\label{BKCxpdyna}
\begin{split}
    \dot{x}_j=\frac{t+\Delta}{2}x_{j-1}- \frac{t-\Delta}{2}x_{j+1},\\
    \dot{p}_j=\frac{t-\Delta}{2}p_{j-1}- \frac{t+\Delta}{2}p_{j+1}.\\
    \end{split}
\end{equation}
显然$x$和$p$在动力学上是完全独立的. 当$t=\Delta$ 时, $x_j$仅受到其左侧近邻$x_{j-1}$的驱动, 而$p_j$仅受到其右侧近邻$p_{j+1}$的驱动, 这种动力学上的非对称性会导致非互易传播的出现. 

\begin{figure}[htp]
\vspace*{2mm}\centering
		\includegraphics[angle=0,width=\linewidth]{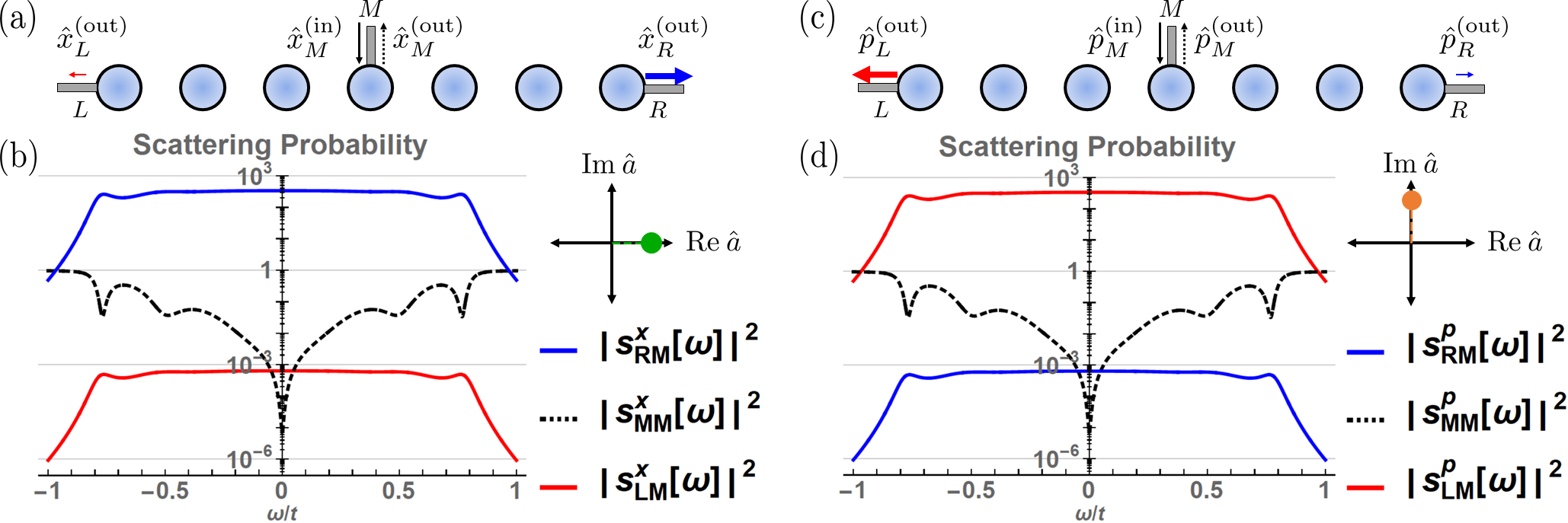}
\bicaption{BKC散射特性. (a) 示意图: 左端、中间、右端三个格点分别耦合到波导, 波导-格点耦合率为$\kappa_L$, $\kappa_M$和$\kappa_R$. 向中间波导输入频率$\omega$的输入信号, 其全局相位$\theta=0$, 对应$x$正交分量激发. 信号向右传播时被放大, 向左传播时被抑制. (b) 散射矩阵元模平方随输入信号频率的变化. 向右(左)传输分量被放大(抑制). 黑色曲线给出反射概率, 其值始终小于等于1. (c)与(a)相同设置, 但输入信号相位改为$\theta=\pi/2$, 对应$p$正交分量激发. (d) 此时信号向左传播被放大, 向右传播被抑制. (b)和(d)的计算参数: 链长$N=13$, 耦合参数强度$\Delta=t/2$, 统一腔内损耗$ \kappa=10^{-2}t$, 波导耦合$\kappa_M=2\kappa_L=2\kappa_R=2t$\cite{PhysRevX.8.041031}. }
{Scattering properties of the BKC. (a) Schematic of the setup. The leftmost, middle, and rightmost
sites are attached to waveguides (coupling rates $\kappa_L$, $\kappa_M$, and $\kappa_R$, respectively). A signal with a frequency $\omega$ and global phase $\theta=0$ corresponding to an $x$ excitation is injected in the middle waveguide and is amplified (deamplified) as it propagates to the right (left).
(b) Amplitude squared of the scattering matrix elements plotted as a function of the frequency of the input signal. As expected, signals
propagating to the right (left) are amplified (deamplified). Note the reflection probability (black) is bounded by unity. (c) Same setup as
in (a), except the phase of the signal is now $\theta=\pi/2$ corresponding to a $p$ excitation. (d) The signal is now amplified (deamplified) as it
propagates to the left (right). For (b),(d), we take $N=13$ sites, $\Delta=t/2$, uniform on-site internal loss rate$ \kappa=10^{-2}t$, and waveguide couplings $\kappa_M=2\kappa_L=2\kappa_R=2t$\cite{PhysRevX.8.041031}.}\label{prxnr}
\vskip 1mm
\end{figure}

如图\ref{prxnr}所示为一条总格点数$N=13$的BKC, 且中间格点(M)、最左端格点(L)和最右端格点(R)均与波导耦合以进行信号输入与输出. 当在中间格点加入一个伴随有相位$\theta$的激发信号, 通过调节$\theta$能够协调输入信号中$x$激发和$p$激发的比例. 如图\ref{prxnr}(a), 当$\theta=0$时输入信号完全为$x$激发, 在M, L, R三个格点均能检测到$x$信号输出, 通过对散射矩阵的计算可以得到信号在不同传播方向上的传输效率. 图\ref{prxnr}(b)反映了当输入信号为$x$激发时, 其在向右传播的过程中被明显放大, 而在左侧波导得到的输出信号被明显抑制. 反之如图~\ref{prxnr}(c, d)所示, 当$\theta=\pi/2$时$p$激发在输入信号中达到完全占比, 此时在输入输出波导检测所得均为$p$信号. 与$x$信号相反, $p$信号在向右传播的过程中被不断抑制, 而在向左传播过程中得到放大. 这一结果证明BKC中能够存在非互易传输, 还提出了一种通过参数调控实现信号放大器或抑制器的新思路. 2024年Andreas团队也于光力腔系统中\cite{2024Optomechanical}成功实现一条长度$N=4$的BKC, 在实验上证明了BKC的非互易传输特性, 也通过对光学微腔衰减速率以及耦合强度的调控实现了信号的放大与抑制.

\section{非厄米拓扑与趋肤效应}
\label{sec4}
非厄米趋肤效应(Non-Hermitian skin effect,NHSE)是非厄米系统中一种奇特的物理效应,它描述了非厄米系统在开放边界条件(open boundary condition, OBC)下本征态波函数局域在系统边界附近的行为, 体现出非厄米哈密顿量的能谱和本征态波函数在周期边界条件(periodic boundary condition, PBC)和OBC下的显著差异\cite{PhysRevLett.121.086803,PhysRevLett.123.066404,PhysRevLett.125.226402,nat.608.50}. 从拓扑能带理论角度分析, NHSE体现的边界敏感性表明在非厄米系统中, PBC下的拓扑不变量不再能准确地描述开放边界条件下系统边界态的性质,证明厄米系统中描述拓扑性质的中心原理“体边对应关系”在非厄米系统中失效\cite{PhysRevLett.123.246801,2020NatPhys.16,PhysRevLett.125.126402}. 因此, 如何基于不同物理系统实现并调控NHSE, 探索并理解其中的新奇拓扑特性,已发展成近年来非厄米物理领域中的热点问题.

\begin{figure}[H]
\vspace*{2mm}\centering
		 \includegraphics[angle=0,width=0.7\linewidth]{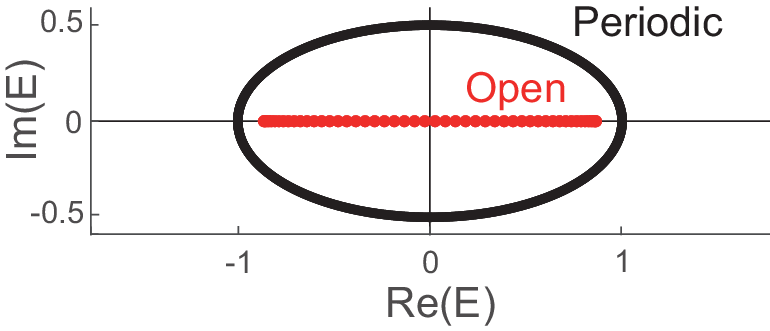}
\bicaption{系统能谱: PBC能谱$E^{\mathrm{P}}_{q,\pm} = t \sin(q)\pm i \Delta \cos (q)$, OBC能谱$E^{\mathrm{O}}_{q} =\sqrt{t^2-\Delta^2}\cos (q)$, 耦合强度参数取$t = 1, \Delta= 0.7$. 对于任意非零$\Delta$, PBC能谱始终为复数, 意味着系统存在参量不稳定性; 相反只要$t>\Delta$, OBC能谱始终为纯实数, 无论系统尺寸如何, 系统均保持稳定\cite{PhysRevB.103.165123}.}
{Spectrum of the system with PBC $E^{\mathrm{P}}_{q,\pm} = t \sin(q)\pm i \Delta \cos (q)$ versus OBC
$E^{\mathrm{O}}_{q} =\sqrt{t^2-\Delta^2}\cos (q)$ with $t = 1, \Delta= 0.7$. The spectrum with PBC is complex for any nonzero $\Delta$, indicating parametric instability. In contrast, the system with OBC is stable as long as $t>\Delta$, regardless of system size\cite{PhysRevB.103.165123}.}\label{BKCenergy}
\vskip 1mm
\end{figure}

Murakami等\cite{PhysRevB.103.165123}研究了BKC模型$H=\sum_{j}  (te^{i \phi_t}a_{j+1}^\dagger a_j +i\Delta  a_{j+1}^\dagger a_j^{\dagger}+\mathrm{H.c.})/2$的动力学矩阵能谱结构, 该模型的动力学演化矩阵的OBC本征值为$E^{\mathrm{O}}_{q} =\sqrt{t^2-\Delta^2}\cos q$, 在$t>\Delta$的范围内其始终为纯实数本征值, 在复平面上表现为一条虚部为零的开放曲线; 而PBC动力学演化矩阵本征值为$E^{\mathrm{P}}_{q,\pm} = t \sin\phi_t \sin q\pm i \sqrt{\Delta^2-t^2 \cos^2\phi_t} \cos q$, 在$\Delta\neq t |\cos \phi_t|$的情况下能谱在复平面上形成闭合环绕结构. 如图\ref{BKCenergy}是$\phi_t =\pi/2$下的复平面能谱, 由于OBC本征值虚部为$0$, 而两种能谱在实部的最大值上具有如下关系:$ t > \sqrt{t^2-\Delta^2}$, 因此PBC能谱必定会相对OBC能谱形成有非零绕数的环绕结构. 

\begin{figure}[H]
\vspace*{2mm}\centering
		 \includegraphics[angle=0,width=0.7\linewidth]{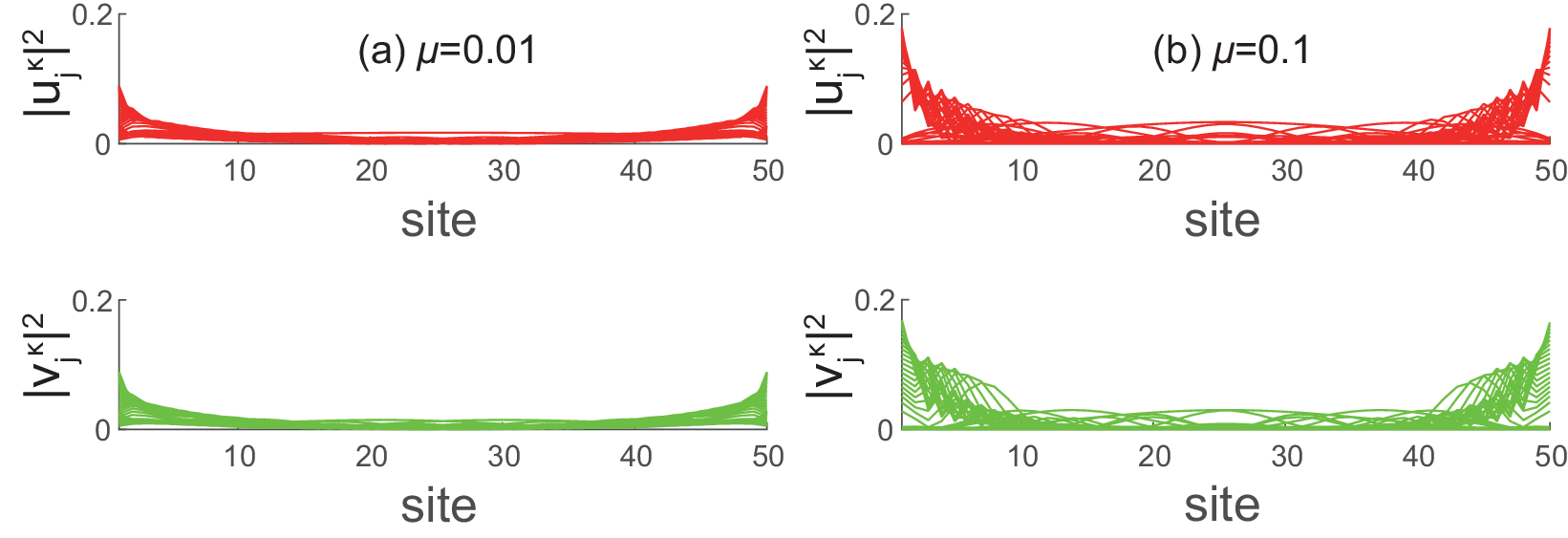}
\bicaption{双边趋肤效应: $N=50$的BKC在不同扰动强度$\mu=0.01(a),\mu=0.1(b)$下的本征态系数向空间两端积累\cite{PhysRevB.103.165123}. }
{Bilocal skin effect in a BKC with $N=50$ under different parameters $\mu=0.01(a),\mu=0.1(b)$. The eigenstate distribution accumulates at the two edges. }\label{NHSE}
\vskip 1mm
\end{figure}

在探究该系统的非厄米趋肤效应时, 形如公式\ref{BKCjuzhen}的实空间非厄米动力学演化矩阵$M$能够通过Bogoliubov基矢变换$(\boldsymbol{a\ a^\dagger})=(\boldsymbol{\alpha\ \alpha^\dagger})T^\dagger$而进行对角化, 其中$\boldsymbol{a}=(a_1,...,a_N),\boldsymbol{\alpha}=(\alpha_1,...,\alpha_N)$:
\begin{equation}\label{duijiaojuzhen}
\begin{split}
   M
    \begin{pmatrix}
		U&V^*\\
		V&U^*\\
	\end{pmatrix}=
    \begin{pmatrix}
		U&V^*\\
		V&U^*\\
	\end{pmatrix}
    \begin{pmatrix}
		\Lambda&0\\
		0&-\Lambda\\
	\end{pmatrix}_{2N},
    \end{split}
\end{equation}
其中$\Lambda$为对角矩阵, 变换矩阵为
\begin{equation}\label{duijiaohua}
\begin{split}
   T=
    \begin{pmatrix}
		U&V^*\\
		V&U^*\\
	\end{pmatrix},
    \end{split}
\end{equation}
对角化后的动力学演化矩阵由两个对角化矩阵$\pm\Lambda$分立描述. 在此基础上为BKC加入扰动项$H_p=\sum_j\mu a_j^\dagger a_j$, 此时两个对角矩阵的PBC复平面能谱会由于扰动项的加入而产生劈裂, 导致两种绕数分别为$W=\pm1$的环绕结构在能谱中同时出现并发生部分相交, 其中相交部分的内部绕数为$W=0$, 而未相交部分的绕数仍然为$W=\pm1$. 这种对角矩阵的特殊能谱现象所导致的非厄米趋肤效应在对角变换后的新基底空间中体现. Bogoliubov基矢变换直接导致新基底向量$\boldsymbol{ \alpha^\dagger}=\boldsymbol{a^\dagger}U-\boldsymbol{a}V$, 其中的任一元素$\alpha^\dagger_k$都是原基底向量的线性组合:
\begin{equation}\label{xinjidi}
\begin{split}
   \alpha^\dagger_k=\sum_j \left( u^k_j a^\dagger_j-v^k_ja_j \right),
    \end{split}
\end{equation}
变换矩阵$T$中的系数$u^k_j,v^k_j$代表了旧基矢在新基矢中的空间分布, 也能反映出系统中是否存在趋肤效应. 如图\ref{NHSE}所示为一条链长$N=50$的一维BKC动力学演化矩阵在经过对角化后的系数$u^k_j,v^k_j$在空间上的分布, 显然系数模平方产生向两侧的对称积累, 这种特征被称为双边趋肤效应. 图\ref{NHSE}(a)所示在扰动较小$(\mu=0.01)$的情况下能谱劈裂较弱, 本征态的双边趋肤效应并不明显. 在逐渐增大扰动$(\mu=0.1)$的情况下[图\ref{NHSE}(b)], 两种绕数结构的劈裂程度增加, 双边趋肤效应的程度增减增强. NHSE的存在对应其PBC能谱出现非零绕数, 在纯虚数耦合$\phi_t=\pi/2$的情况下环绕结构的必要条件是TMS相互作用系数$\Delta\neq0$. 一旦$\Delta=0$, PBC能谱会从闭合环结构退化为虚部为零的开放曲线, NHSE也会随之消失, 这证明压缩相互作用的引入为厄米系统带来了非厄米动力学特性.

\begin{figure}[H]
\vspace*{2mm}\centering
	\includegraphics[angle=0,width=0.5\linewidth]{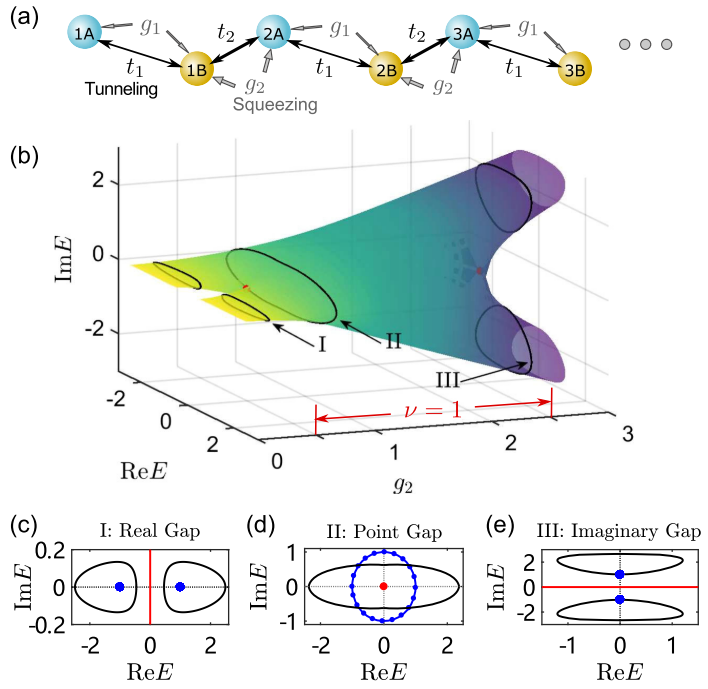}
\bicaption{(a) TMS存在下由A, B类格点构成的压缩SSH模型. 相邻格点间的BS强度与TMS强度分别记为$t_1, t_2$和$g_1, g_2$. (b) 在$t_2=3t_1/2$且$g_1=0$情况下复平面中的能谱随压缩强度$g_2$的变化. $g_2 = 0.5t_1, 2.5t_1$处的红点标记$E=0$处点隙拓扑闭合和打开的临界点; 拓扑指数$\nu=1$对应$g_2 \in (0.5, 2.5)t_1$区间. (c) - (e) (b)中标记为 I, II, III的能谱(黑色曲线)可连续形变至: (c) $\pm1$(蓝色点), (d) 单位圆 (蓝色圆), (e) $\pm i$ (蓝色点), 同时保留相应的能隙(红色)\cite{PhysRevLett.130.203605}.}
{(a) Schematic of the squeezed SSH model consisting of the A and B sublattices in the presence of TMS. The BS and TMS strengths between the adjacent sites are denoted by $t_1, t_2$ and $g_1, g_2$, respectively. (b) Spectrum in the complex plane as varying the intercell squeezing strength $g_2$. Here $t_2=3t_1/2$ and $g_1=0$. The red dots at $g_2 = 0.5t_1, 2.5t_1$ represent the critical points for closing or opening the point gap at $E = 0 $ and $ ν =1 $ corresponds to $g_2 \in (0.5, 2.5)t_1$. (c) - (e) Spectra (black curves) for I, II, and III in (b) can be continuously deformed to (c) $\pm1$ (blue dots), (d) unit circle (blue circle), and (e) $\pm i$ (blue dots), respectively, while preserving the associated gaps (red)\cite{PhysRevLett.130.203605}.}\label{SSH}
\vskip 1mm
\end{figure}

在BKC的基础上对模型进行改造会诱导出更丰富的非厄米拓扑特性. 如图\ref{SSH}(a)所示, 2023年吕新友等\cite{PhysRevLett.130.203605}在玻色SSH模型的基础上, 向单元内部格点间、不同单元间引入强度不同的TMS相互作用, 得到二次型玻色SSH模型:
\begin{equation}\label{SSHHam}
\begin{split}
    H_{\mathrm{SSH}}=&\sum_j (t_1 a_{j,A}^\dagger a_{j,B} +t_2 a_{j+1,A}^\dagger a_{j,B}+g_1 a_{j,A} a_{j,B}+g_2 a_{j+1,A} a_{j,B}+\mathrm{H.c.}).
    \end{split}
\end{equation}

在经过同样的傅里叶变换后, 该系统的动量空间动力学矩阵有两个双重简并的本征值$E_\pm(k)=\Delta^2+2(t_1t_2-g_1g_2)\cos k \pm2i(t_1g_2-t_2g_1)\sin k$, 其中$\Delta=\sqrt{t_1^2+t_2^2-g_1^2-g_2^2}$. 如图\ref{SSH}(b)所示, 当仅保留单元间的TMS相互作用即$g_1=0, g_2\neq0$后, 在$t_2=3t_1/2$的参数空间下压缩SSH模型的动量空间动力学矩阵能谱随$g_2$展现出连续拓扑相变, 包括三个能隙闭合或张开的临界点以及三种拓扑结构. 当$g_2=0$时, 系统中不存在TMS相互作用, 能谱收缩为两条不相连的开放线条, 体系不具备非厄米动力学特征. 当$g_2\in(0,0.5)t_1$时, 由于压缩相互作用的引入为本征值带来非零虚部, 导致能谱出现两个被实数隙$\text{Re}E=0$分隔的孤立环状谱[图\ref{SSH}(c)]. 当$g_2>2.5t_1$时, 能谱依然为两个孤立环状结构, 但能隙由实数隙转变为纯虚数隙$\text{Im}E=0$[图\ref{SSH}(e)]. 在$g_2=0.5t_1,2.5t_1$处, 能谱发生能隙闭合, 此时两个孤立的环状谱合并为一条绕能量原点$E=0$具有稳定绕数的对称环形谱, 即产生了稳定的点隙拓扑结构[图\ref{SSH}(d)], 此时动力学矩阵的Kramers态出现双边趋肤效应.

 \section{非厄米动力学与量子关联}
在量子光学领域, SMS\cite{PhysRevA.13.2226,PhysRevD.23.1693,PhysRevLett.55.2409}和TMS作用\cite{PhysRevA.31.3068,PhysRevLett.59.2153}不仅被用来降低光场正交分量的量子噪声, 还被广泛应用于量子纠缠、关联的制备以及量子隐形传态等\cite{science.282.5389.706}. 因此含有压缩作用的QBS为研究非厄米动力学与量子关联等效应的内在联系、实现基于非厄米特性的量子调控提供了重要平台.

\begin{figure}[H]
\vspace*{2mm}\centering
	\includegraphics[angle=0,width=0.5\linewidth]{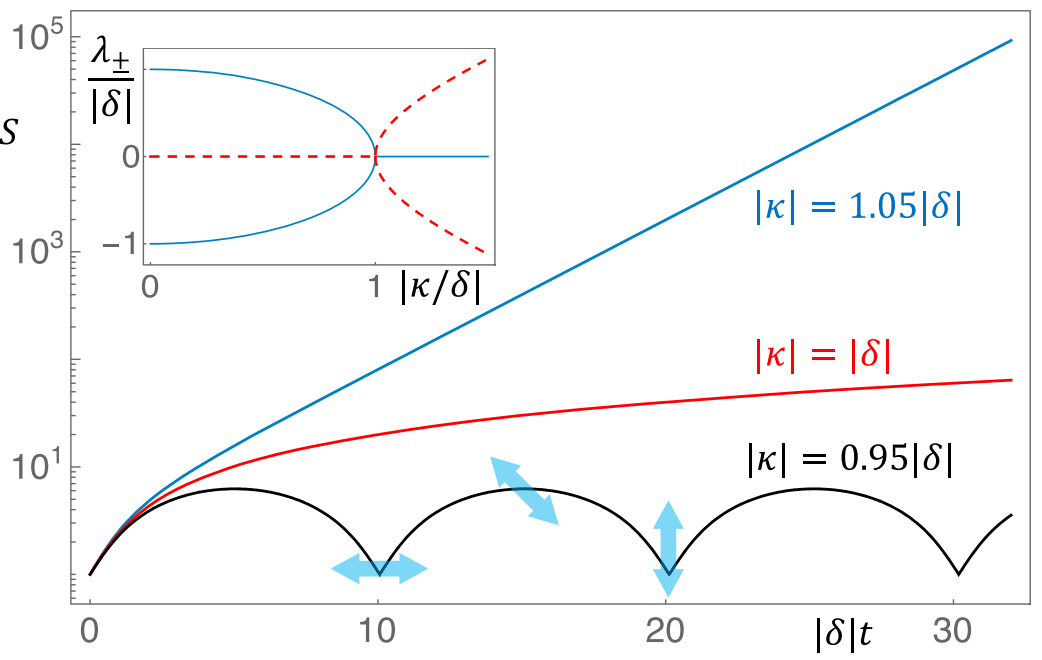}
\bicaption{由EP诱导的量子压缩动力学. (a) 不同$\kappa$值$|\kappa/\delta|= 0.95, 1$和$1.05$下压缩因子随演化时间的变化. 蓝色箭头指示$\kappa, \delta > 0$时的TMS方向. 插图给出与图\ref{eigen}相同的本征值\cite{PhysRevLett.128.173602}.}
{EP induced quantum squeezing dynamics. (a) The squeezing factor versus evolution time for different $\kappa$: $|\kappa/\delta|= 0.95, 1,$ and $1.05$. Blue arrows indicate the two-mode squeezing directions for $\kappa, \delta > 0$. The inset shows the eigenvalues, which is same as Fig. \ref{eigen}\cite{PhysRevLett.128.173602}.}\label{yasuo}
\vskip 1mm
\end{figure}

2022年, 杜胜望等\cite{PhysRevLett.128.173602}在两模QBS中发现压缩因子在非厄米动力学演化矩阵的EP点附近出现动力学行为改变, 考虑系统$ H=\delta(a_1^\dagger a_1 + a_2 ^\dagger a_2)+i\kappa(a_1^\dagger a_2^\dagger - a_1 a_2)$, 其非厄米动力学演化矩阵有二重简并本征值$ \lambda_\pm =\pm \sqrt{\delta^2-\kappa^2}$以及在$|\delta|=|\kappa|$处的二阶EP. 通过海森堡动力学方程能够得到产生湮灭算符随时间的演化为:
\begin{equation}\label{}
\begin{split}
    \begin{pmatrix}
		a_1(t)\\
		a_2^\dagger(t)\\
	\end{pmatrix}
    =\begin{pmatrix}
		A&B\\
		B^*&A^*\\
	\end{pmatrix}
    \begin{pmatrix}
		a_1(0)\\
		a_2^\dagger(0)\\
	\end{pmatrix},
    \end{split}
\end{equation}
其中
\begin{equation}\label{}
\begin{split}
   A=\sum_{s=\pm}\frac{(\lambda_s-\delta)e^{i\lambda_s t}}{2\lambda_s},
    B=\sum_{s=\pm}\frac{\kappa e^{i\lambda_s t}}{2i \lambda_s}.
    \end{split}
\end{equation}
量子压缩因子$S=|A|+|B|$是对系统压缩程度的量化. 如图\ref{yasuo}所示, 在EP点附近压缩因子动力学演化行为发生明显改变. 当$|\kappa|=1.05|\delta|$时, 压缩因子呈指数级动态增长:$S\backsimeq(\kappa/\sqrt{\kappa^2-\delta^2}) e^{\sqrt{\kappa^2-\delta^2}t}$. 当$|\kappa|=0.95|\delta|$时, 压缩因子呈周期性振荡演化, 其中振荡周期为$T=\pi/\sqrt{\kappa^2-\delta^2}$, 振荡峰值为$S_{\mathrm{max}}=\sqrt{(\delta+\kappa)/(\delta-\kappa)}$. 当系统处于动力学矩阵EP点$|\delta|=|\kappa|$时, 压缩因子$S=\sqrt{1+\delta^2 t^2}+\kappa t$会在长时间尺度$\delta t\geqslant1$呈线性增长. 

2025年, 何琼毅、黄馨瑶等\cite{yu2025}在引入单模压缩的BKC模型$ H=\sum_{j=1}^{N-1} \left(g a_{j}^\dagger a_{j+1} +J a_{j}^\dagger a_{j+1}^{\dagger}\right) +\sum_{j=1}^N \eta (a_j^\dagger)^2/2+\mathrm{H.c.}$中提出基于非厄米EP的多模式纠缠动力学调控方案. 将多模玻色链的动力学演化映射至BdG框架后得到不同阶数的EP, 且EP划分的谱区域精确对应三类纠缠动力学. 以两模系统为例, SMS使$g=J$处的两个二阶EP劈裂, 从而将能谱相图分割为三种区域, 分别对应纯虚数、纯实数和复数本征值. 模式间的纠缠动力学演化则在三个区域分别呈指数增加, 振荡以及指数增加与振荡叠加的行为. 他们还证实通过引入非整数π相位或调控相互作用强度不相等可以实现多模系统中的高阶EP, 并发现相较于二阶EP, 高阶EP可显著增强多模纠缠程度, 显示出该方案在纠缠调控中的独特优势. 

此外, 最近的相关研究还包括: 在BKC中实现纠缠相变\cite{PRXQuantum.5.010313}、通过拓扑相建立稳态纠缠\cite{PhysRevA.108.062405}、通过高阶EP增强量子Fisher信息\cite{fisherEP}和在SSH模型中引入TMS实现量子传感\cite{PhysRevResearch.7.013309}等. 这些研究均以QBS作为载体, 通过引入压缩作用, 探索非厄米动力学与量子效应的内在联系, 基于非厄米特性实现对量子效应的有效调控. 

\section{结论与展望}
本文综述了以分束器型相互作用与双模压缩相互作用为核心的二次型玻色系统中非厄米动力学的研究进展. 文中主要回顾了从两模到多模拓展的二次型玻色系统(包括玻色Kitaev链、玻色SSH模型等), 在BdG框架下通过构造等效非厄米动力学所呈现的丰富非厄米特性, 具体涵盖奇异点、正交性非互易、非厄米拓扑及趋肤效应等新奇物理现象. 值得注意的是, 二次型玻色系统中的双模压缩作用可同步诱导量子关联效应, 这一特性使该系统成为深入探究非厄米动力学与量子关联内在联系的天然载体. 基于此, 本文亦介绍了近期利用非厄米动力学调控系统量子关联的相关研究成果, 为该领域的后续研究梳理脉络. 

在实验上, 量子技术尤其是玻色平台的迅速发展为实现二次型玻色系统提供了多种途径. 基于光子与机械振子的玻色子特性, 目前可行的实验平台包括光力系统\cite{natphy2023,2024Optomechanical,2022nat41586}、集成光学芯片\cite{2025nat329,2025light}、多模机械振子与超导量子比特耦合系统\cite{2024natphys20,2024natphys1448}、以及超导量子线路\cite{2024Quantum}等, 通过在上述量子系统采取有效调控机制, 都可以实现系统中模式间的分束器型和压缩相互作用, 为构建等效的二次型玻色系统并进一步观测其中新奇的非厄米特性提供实验验证.

最后, 二次型玻色系统作为免受量子跳跃影响的无耗散量子体系, 不仅能高效实现多种非厄米动力学, 更为基于非厄米特性的量子效应调控提供了重要平台. 现有研究多聚焦于少体体系与双模压缩作用, 因此未来的核心研究方向可围绕以下方面展开: 如在二次型玻色系统中进一步引入相位调控、单模压缩及模式间长程耦合等机制, 探究其诱导的新型非厄米动力学与新奇拓扑现象, 挖掘非厄米特性与多种类量子效应的深层联系并阐明其物理机理等. 上述研究旨为基于非厄米特性的量子调控提供全新思路, 推动非厄米物理与量子物理交叉领域理论与应用的深入发展.

\newpage

    \maketitle
    \abstract[english]{
        Non-Hermitian physics has emerged as a rapidly advancing field of research, revealing a range of novel phenomena and potential applications. Traditional non-Hermitian Hamiltonians are typically simulated by constructing asymmetric couplings or by introducing dissipation and gain to realize non-Hermitian systems. The quadratic bosonic system (QBS) with squeezing interaction is intrinsically Hermitian; however, its dynamical evolution matrix in both real and momentum spaces is non-Hermitian. Starting from the Heisenberg equation in real space, the dynamical evolution matrix of the system features higher-order non-Hermitian exceptional points. Based on this, applying a field-operator transformation $\{\hat{x},\hat{p}\}$ to the dynamical evolution matrix yields quadrature nonreciprocal transmission between the $\hat{x}$ and $\hat{p}$ operators. This nonreciprocal characteristic can be utilized in signal amplifiers. On the other hand, within the Bogoliubov–de Gennes framework in momentum space, one can observe non-Hermitian topological phenomena such as point-gap topology and the non-Hermitian skin effect, both induced by spectra with nonzero winding numbers. Additionally, QBS can be employed to realize non-Hermitian Aharonov–Bohm cages and to extend non-Bloch band theory. Previous studies in non-Hermitian physics have largely concentrated on classical systems. The influence of non-Hermitian properties on quantum effects remains a key issue awaiting exploration and has evolved into a research direction at the interface of non-Hermitian and quantum physics. In QBS, squeezing interactions without dissipation cause the dynamical evolution of the system to display effective non-Hermitian characteristics and induce quantum correlation effects. Recent studies have shown that the vicinity of non-Hermitian exceptional points in QBS can alter squeezing dynamics and entanglement dynamics. Therefore, such systems not only offer a natural platform for realizing quantum non-Hermitian dynamics but also constitute an important basis for investigating the relationship between non-Hermitian dynamics and quantum effects, as well as for achieving quantum control based on non-Hermitian properties. Future research may further focus on elucidating the connections between non-Hermitian dynamics and quantum effects in QBS, which is expected to serve as a bridge linking non-Hermitian dynamics and quantum effects.}
    \keywords[english]{
        non-Hermitian dynamics, quadratic bosonic systems, nonreciprocity, non-Hermitian skin effect, quantum correlations
    }
    \fund[english=true]{Project supported by the National Natural Science Foundation of China (Grants No. 12474353 and No. 12474354), the Aviation Science Foundation of China (Grant No. 20240058051004), and the Fundamental Research Funds for the Central Universities.}

    \bibreference[
        nocite, 
        newpage
    ]{Ref}

\end{document}